# Time-of-Flight Photoelectron Momentum Microscopy at 100-500 MHz Synchrotron Sources: Electron-Optical Chopping or Bandwidth Pre-Selection

G. Schönhense, K. Medjanik, O. Fedchenko, A. Zymakova, S. Chernov,

D. Vasilyev, S. Babenkov, H. J. Elmers

*Institut für Physik, Johannes Gutenberg-Universität, 55128 Mainz, Germany*

P. Baumgärtel, P. Goslawski

*BESSY II, Helmholtz-Zentrum, 12489 Berlin, Germany*

G. Öhrwall

*MAX IV Laboratory, Lund University, P.O. Box 118, SE-221 00 Lund, Sweden*

M. Ellguth, and A. Oelsner

*Surface Concept GmbH, 55124 Mainz, Germany*

**ABSTRACT**

The small time gap of synchrotron radiation in conventional multi-bunch mode (100-500MHz) is prohibitive for time-of-flight (ToF) based electron spectroscopy. Even the new generation of delay-line detectors with improved time resolution (<100ps) yields only 20-100 resolved time slices within a 2-10ns gap. Here we present two techniques of implementing efficient ToF recording at sources with high repetition rate. A fast electron-optical beam blanking unit with GHz bandwidth, integrated in a photoelectron momentum microscope, allows chopping the photon-pulse train to any desired repetition period. Aberration-free momentum distributions have been recorded at 'chopped' pulse periods of 5MHz (at MAX II) and 1.25MHz (at BESSY II). The approach is benchmarked against the alternative way of implementing a dispersive element, e.g. a hemispherical analyzer, in the electron optics. Both approaches, chopping in the time domain as well as bandpass pre-selection in the energy domain, can enable efficient ToF spectroscopy, spectroscopic real-space imaging and momentum microscopy with few-meV resolution using 100-500MHz Synchrotron radiation, highly-repetitive lasers or cavity-enhanced high-harmonic sources. For comparison, we show results recorded at BESSY II with a 'parasitic' 4-bunch island-orbit pulse train, coexisting with the 500MHz filling pattern on the main orbit.

# I. Introduction

Thanks to its perfectly-periodic time structure, synchrotron radiation is an ideal tool for time-of-flight (ToF) photoelectron spectroscopy [1], real-space spectroscopic photoemission electron microscopy (ToF-PEEM) [2,3], angle-resolved ToF spectroscopy [4-6] or momentum microscopy (MM) [7]. Covering a huge interval in ($E_{kin}$,**k**) parameter space in a single exposure, the ToF-MM approach has proven powerful in the VUV [8], soft-X-ray [9] and hard-X-ray range [10,11]. All these ToF-based methods establish a highly-effective approach to angular-resolved photoelectron spectroscopy (ARPES) [12-14], the method of choice for studies of the electronic structure. This branch of photoemission is steadily growing, being fueled by the discovery of topological states and exciting electronic properties of quantum materials [15-17].

Recently, the advent of full-field-imaging MM as new approach to hard-X-ray photoelectron diffraction (hXPD) [18] has opened an avenue toward fast and effective recording of structural information. The high site specificity of hXPD [19] can be exploited as unique fingerprint of atomic sites in compounds. The first applications of this young technique for the analysis of Mn dopant sites in the diluted ferromagnetic semiconductor (GaMnIn)As [20] and Te in the Si lattice in the hyperdoped regime [21] shine a first light on the potential of hXPD. Electronic and structural information can be obtained in a single experiment and at identical conditions (kinetic energy of the photoelectrons, size and position of the probing spot, probing depth). In turn, this allows eliminating the strong hXPD patterns imprinted on the valence-band momentum images in hard X-ray ARPES (HARPES), allowing mapping of intensity-corrected valence bands [22,23,11,20]. A time-resolved variant of ToF-MM with fs-pulsed sources was recently established at the free electron laser FLASH at DESY, Hamburg [24-26].

The efficiency of the three-dimensional recording architectures of angular-resolved ToF spectrometers and MMs is defined by the time-resolving image detector. Delay-line detectors (DLDs) [27,28] or the upcoming solid-state pixel arrays [29,30] are characterized by time resolutions in the 50-200ps range. Hence, for ToF-based experiments a pulse period in the range of 50–500ns would be favorable, in order to provide a sufficiently long time gap for resolving ~$10^3$ time slices in a spectrum. The vast majority of experiments at synchrotron-radiation sources requires maximum photon flux and do not need time structure. Hence, most synchrotron run time provides users quasi-CW X-rays with pulse rates between 100 and 500MHz, with limited time dedicated to "few-bunch" modes for experiments that exploit the pulsed nature of the light. Corresponding periods of 2-10ns are too short for high-resolution ToF-electron spectroscopy with a sufficient number of resolved times slices. The same arguments apply to the upcoming photon sources based on high harmonic generation (HHG) with intra-cavity conversion [31-33] and for UV laser sources running at pulse rates in the 80MHz range.

Up to now there are basically two approaches for the selection of longer periods between the photon pulses. The first approach is realized as an electron bunch separation scheme, where a single bunch or few bunches are displaced in the transversal plane from the main orbit and filling pattern by different techniques. At the ALS, Berkeley, a high-repetition-rate vertical kicker magnet is used to displace one single bunch of the filling pattern to a different vertical orbit [34]. At BESSY II, one single bunch of the filling pattern is blown up in the transverse dimensions by pulse picking resonant excitation [35]. An aperture in the beamline blocks the radiation pulses from the main orbit, only selecting the blown up parts of the excited bunch, resulting in a reduced repetition rate, but with reduced intensity. Kicking the electrons with a selectable angle through an insertion device is also an important ingredient in femtosecond-slicing experiments [36]. At BESSY II, another technique is under development, which is based on a special storage ring setting, generating a second stable 'island orbit', wiggling around the main orbit in the equatorial plane [37]. All these transverse bunch separation schemes provide two different radiation sources in bending magnets and undulators for the beamlines, which could be selected by apertures.

The second approach is a fast mechanical chopper for the radiation pulses within the beamline on the basis of a high-speed rotating disc with slits. The rotation frequency is synchronized with the storage-ring period which requires precise phase matching with the filling pattern of the storage ring. Pioneering work has been performed at the ESRF (Grenoble, France) [38] and BESSY II (Berlin, Germany) [39]. In the latter experiment the so-called camshaft bunch (single bunch in a gap of the filling pattern) has been successfully used for angular-resolved ToF spectroscopy. This camshaft bunch is separated by a gap of 60ns on both sides from the multibunch train, where the pulses have only 2ns spacing. Mechanical chopping is highly demanding in terms of stability and synchronization. The chopper at BESSY II operates at a frequency of 1.25MHz.

In this article we describe two novel methods that do not require an electron-bunch separation within the storage ring. In the first approach the 'photon-pulse selection' is performed by electron-optical means on the experimental side, using a special time-of-flight MM. In a photoemission experiment the photon pulses are converted into a pulsed electron signal with the same time structure. The method employs a fast electric beam deflector with Gigahertz bandwidth integrated between two pinholes in the imaging column of a photoelectron MM. The time-resolving image detector (DLD) is used for spatio-temporal beam diagnostics and for time-of-flight k-space and real-space imaging. Results are shown from two different experiments: Photoelectron pulses generated by a soft-X-ray pulse train with 200ns period (chopper frequency 5MHz) have been selected out of the 100MHz multibunch train at the (former) storage ring MAX II (Lund, Sweden) [40]. Electron pulses excited by a vacuum ultraviolet (VUV) beam with 800ns period have been selected in the so-called 'hybrid mode' of BESSY II (Berlin). Full extinction of all undesired photoelectron bunches is reached by proper positioning of two beam crossovers in two pinholes in the electron path. When passing the fine pinholes, the momentum image is encoded in terms of the angular distribution of the electrons (reciprocal image). The performance of the experiment using the chopped camshaft pulse is compared with the conditions for a special 5MHz island-orbit filling pattern of BESSY II.

An alternative way of chopping the electron pulse train in the time domain by a GHz deflector is pre-selection of a reduced bandpass in the energy domain. This confines the width of the electron spectrum entering the ToF analyzer, so that the full pulse rate can be retained. We studied this approach using a hemispherical analyzer as dispersive element in a setup with an 80MHz UV-laser source. The first prototype of the 'dispersive-plus-ToF' hybrid instrument will be installed at DIAMOND (beamline I09). Both methods can enable efficient ToF spectroscopy, real-space imaging and momentum microscopy with few-meV resolution at 100-500MHz Synchrotron radiation sources, highly-repetitive lasers or cavity-enhanced high-harmonic sources.

## 2. Experimental technique and results
### 2.1 Layout of the fast electron-optical chopper

The measurements were carried out using soft X-ray pulses at beamline I 1011 [41] of MAX II (Lund, Sweden) and VUV pulses at beamline U125-2_NIM [42] of BESSY II (Berlin) in normal multibunch operation. For comparison, a reference measurement was performed at BESSY II using a special 'island orbit' with a different filling pattern of 4 bunches, oscillating around the equatorial plane of the storage ring. This orbit is spatially separated from the common multibunch train; the synchrotron radiation from this few-bunch pulse train is selected by inserting an aperture at an intermediate focus of the beamline.

Beamline I 1011 at MAX II provided linearly, circularly, and elliptically polarized X-rays in the range 0.2-1.7keV. The source was an elliptically-polarizing undulator of the APPLE II type with a period of 46.6mm, installed on the 1.5GeV MAX II storage ring, and the beamline was a collimated PGM design [41]. For the experiments presented here, linearly polarized photons with an energy of 500eV were used, with an exit-slit size of 100μm giving a FWHM of ~200meV at a photon energy of 500eV.

Beamline U125-2_NIM at BESSY II comprises a 10m normal-incidence monochromator providing linear polarized photons in the energy range of 6 to 35eV. Energy bandwidths of less than 10meV can easily be achieved with the 300 l/mm grating and entrance and exit slit sizes of 20μm in the first order. Spectral impurities due to higher harmonics are effectively suppressed in the U125-2 undulator by its quasi-periodic structure of the magnets.

Fig. 1 shows the electron-optical scheme of the experiment. The high-frequency (HF) chopper is implemented into the column of a ToF MM. This special cathode-lens type electron microscope is optimized for best resolution in *k*-space and parallel acquisition of an energy interval of several eV width; for details, see [9]. Momentum microscopes make use of the fact that a reciprocal image with high *k*-resolution and a linear $k_\parallel$-scale occurs in the Fourier plane (backfocal plane) of the objective lens. The electron lenses in the microscope column project the magnified momentum image via a low-energy drift space for energy dispersion to the detector plane. The electron distribution in this plane is recorded by a 3D-resolving DLD [27,28]. The 'bunch selector' in the electron-optical column consists of *two pinholes* in two additional beam crossovers and the *HF deflector* placed in between. In the present experiment a simple quadrupole deflector was used, higher time resolution can be achieved with more sophisticated electrode arrangements. The deflector is supplied by coaxial lines with bandwidth 20GHz (via SMB vacuum feedthroughs) and 50 Ohm terminations. Fast voltage pulses with amplitudes up to 20V are superimposed to the static potential of the deflector plates, centering the beam. The kinetic energy in the deflector is ~100eV. The pulse generator is phase-locked to the clock of the pulsed photon beam, e.g. the bunch marker of the storage ring. Pinholes 1 and 2 are piezomotor-adjustable arrays of size-selectable apertures, located in the planes of the beam crossovers.

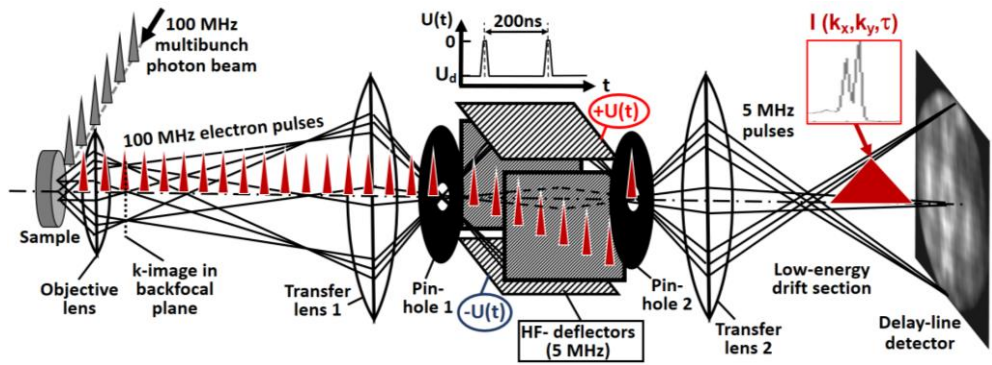

**Figure 1**
Schematic view of the bunch selector in a time-of-flight momentum microscope operated at full filling pattern of a storage ring. The depicted example corresponds to the situation at MAX II (100MHz bunch-filling pattern). The electron-optical system comprises three lens groups which transport and magnify the momentum image from the backfocal plane of the objective lens (left) to the delay-line detector (right). The sample emits a photoelectron pulse train of 100MHz and each of the pulses (indicated by red triangles) contains the full spectrum. The high-frequency (HF) deflector is located between two narrow beam crossovers (pinholes 1 and 2). The deflector voltage *U(t)* (inset on top) deflects the electrons originating from the multibunch train, except for a short time interval where the voltage is zero, thus transmitting the electrons from the desired individual pulse through the second aperture (pinhole 2). This way, a 5MHz sub-period of pulses is generated, which is dispersed in time-of-flight τ the low-energy ToF section. The radial coordinate is strongly exaggerated, the screen shows a W4f$_{7/2}$ photoelectron diffraction pattern recorded at hν=500eV. Note that each red triangle represents the full 3D array I(k$_x$,k$_y$,τ); in the pinholes the momentum pattern is encoded as angular distribution.

The static and dynamic parts of the deflector voltages are adjusted such that the selected electron bunches with the desired period are transmitted through *pinhole 2*. All other electron bunches are blocked by deflecting them off-axis (cf. Fig. 1). The proper setting of the time interval is monitored by removing pinhole 2 and adjusting transfer lens 2 such that a real-space image of the plane of pinhole 2 is visible at the DLD at the end of the column [examples in Figs. 2(a) and 3(b)].

## 2.2 Chopper operation with 5 MHz at MAX II (100 MHz filling pattern)

As an example, Fig. 2 shows a measurement of the tungsten 4*f* core-level and valence-band spectra, recorded at a photon energy of 500eV at beamline I 1011 of MAX II. Fig. 2(a) shows an I(y,t) cut through the 3D spatiotemporal intensity distribution on the detector, recorded for fully-opened *pinhole 2*. The pulse generator as well as the DLD (start signal) are phase-locked to the bunch marker of the storage ring. The two parallel lines in (a) represent the W4*f* spin-orbit doublet, dispersed in the ToF section. This doublet shows up many times in the pattern, with spacing 10ns, corresponding to the photon pulse rate of 100MHz from the storage ring. It is eye-catching that one 4*f* doublet is located on the electron-optical axis (chain line), whereas all others are displaced off center by the static part of the deflector voltage $U_d$. They are truncated by the rim of the image detector at the lower end of the panel. For this measurement, *transfer lens 2* was adjusted so that a real-space image is focused on the detector in order to visualize the 2D distribution.

*Pinhole 2* has been moved stepwise into the beam as sketched in Fig. 2(a), positions (b, c and d). In spectrum (b), taken with off-center pinhole, we observe the full multibunch train of MAX II with a period of 10 ns. The drop toward the rims of the time window is an artefact due to the trigger mode of the DLD. Moving the selector aperture to the next position yields a spectrum with partial extinction (c), and position (d) yields almost total extinction, with a small residual intensity from the other bunches. Changing transfer lens 2 to the momentum-imaging mode and setting a high drift energy leads to a wide-range spectrum (e) with full extinction of undesired signals.

The temporal dispersion in the electron-optical column upstream of *pinhole 1* is very small and hence negligible. The HF chopper operates achromatic in a large energy range of several 100eV width. Figure 2(e) shows the complete photoemission spectrum from the low-energy cutoff ($E_{kin}$= 0) at 118ns via the W 4*f* signal at 60ns ($E_{kin}$= 460eV) to the Fermi-energy cutoff at 22ns ($E_{kin}$= 495eV), recorded at fixed settings in a single exposure. The temporal dispersion dτ of the ToF drift section of length $L_d$ depends strongly on the drift energy $E_d$:

$$\frac{d\tau}{dE_d} = -\frac{L_d}{2\sqrt{2/m_e}} \frac{1}{E_d^{3/2}}. \qquad (1)$$

Due to the large variation of the drift energy between the low-energy cutoff at $E_d$=30eV and the Fermi cut-off at $E_d$=525eV the temporal resolution varies from 31ps/eV to 2.2ns/eV, respectively. Given the time resolution of the DLD (150ps), the 4*f* doublet is not resolved in the survey spectrum, Fig. 2(e), but is well resolved when recorded at lower drift energy, Figs. 2(f,g).

In the k-imaging mode the chromatic aberration of the lens system limits the depth of focus to an energy band of up to 10eV, depending on the size of the field aperture, which acts as a contrast aperture for k-imaging. It is a special property of the momentum microscope that different energy regions in a spectrum can be selected to be in focus by tuning just one parameter, namely the sample bias. The width of the spectrum can be restricted to a region of less than 10eV below $E_F$ by operating a section of the optics as high-pass filter (for details, see [43]). However, this comes at the expense of the depth of focus because in the region of the saddle point the lens aberrations are very large.

The valence-band and core-level spectra Fig. 2(f-h) have been recorded on an expanded time-of-flight scale by selecting the proper drift energies in the ToF section. The valence-band spectrum is rather noisy since it suffers from a large background originating from higher-order contributions in the photon beam. The much stronger W 4f signal is very pronounced. Part of the surface is covered by chemisorbed oxygen, as visible in the well-known chemical shift of the W 4f peaks in Fig. 2(g).

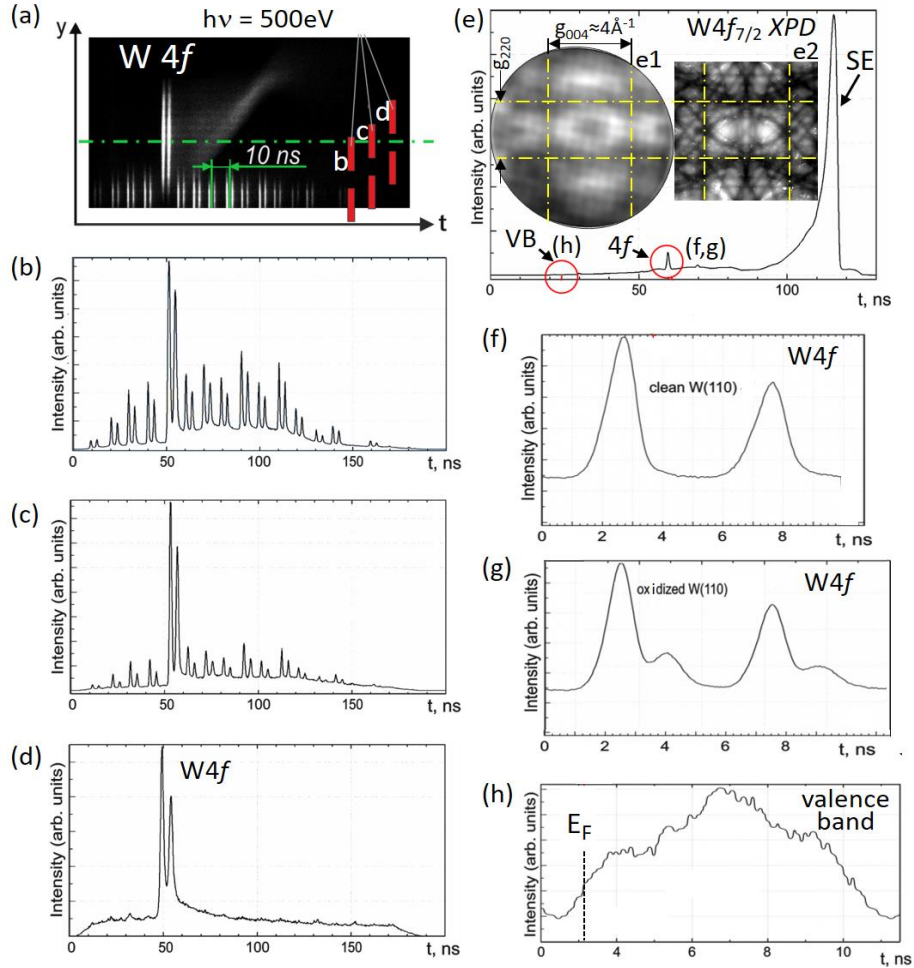

**Figure 2**
Operation of the *HF deflector* at 5MHz at beamline I 1011 at MAX II (filling pattern 100MHz); data recorded for a W(110) surface at hν=500eV. (a) Spatiotemporal action visible in a measured *I(y,t)* cut through the 3D intensity distribution on the delay-line detector for fully-opened *pinhole 2*. The two closely-spaced lines correspond to the W $4f_{7/2,5/2}$ doublet, shifted along y by the fast deflector. (b,c,d) *I(t)* spectra with different degree of extinction, taken for three different positions of the selector aperture *pinhole 2* as indicated in (a). (e) Survey ToF spectrum (range ~500eV) recorded for complete extinction; spectrum strongly dominated by the secondary-electron peak (SE). Inset (e1), measured X-ray photoelectron diffraction XPD pattern for W $4f_{7/2}$ with the main Kikuchi bands being marked by dashed lines; (e2) calculated W $4f_{7/2}$ XPD pattern. (f,g) W 4f core-level doublet and (h) valence-band spectrum taken at lower drift energy of 30eV shown on a stretched time scale. The high background in the valence-band region (h) originates from higher-order contributions of the monochromator. Parts of the surface showed an oxidic overlayer (g).

The W 4f core-level signal can be used for full-field imaging of the X-ray photoelectron diffraction (XPD) pattern [44]. When the electrons are crossing the small selector aperture (*pinhole 2*), the XPD pattern is encoded in terms of their angular distribution. As an example the W $4f_{7/2}$ XPD pattern of the W(110) surface is shown as inset (e1) in Fig. 2. The characteristic diffraction pattern is clearly visible without significant distortion in the 5MHz chopping mode. The finite size of the selector aperture does not influence the resolution in the XPD image because for the angle-encoded diffraction pattern a small parallel shift of the beam inside of the aperture (~100μm) is insignificant for the momentum image on the DLD (40mm diameter). A calculated W 4*f* XPD pattern for W(110) is shown in inset (e2); for details on the Bloch-wave model, see [18-20]. Although the calculated pattern shows many more details, the similarity of measured and calculated patterns allows assigning the principal Kikuchi bands corresponding to reciprocal lattice vectors $g_{004}$ (vertical band) and $g_{220}$ (horizontal band) as marked in Fig. 2(e).

### 2.3 Selection of the 'camshaft' pulse with 1.25 MHz at BESSY II (500 MHz filling)

The pulse-selecting approach can be used not only for ToF spectroscopy, real-space imaging and photoelectron diffraction but also for *momentum imaging of valence bands*. This application was explored at the storage ring BESSY II in Berlin using the 'hybrid' filling pattern, consisting of a 500MHz multibunch train plus a solitaire pulse (usually with higher bunch charge) in a gap of ~120ns width. The separation of this 'camshaft' pulse from the multibunch train is less demanding than the complete extinction of the neighboring bunches in the 100MHz pulse train of MAX II. The relatively large time gap on both sides of the solitaire pulse relaxes the requirement for the slope of the leading and trailing edges of the blanking signal. Electrostatic gating of the camshaft pulse at BESSY II has previously been shown using a blanker electrode in a PEEM [45] and gating of the MCP-detector was applied in an angle-resolved ToF spectrometer [46]. In the present experiment we used a rectangular blanking signal U(t) of ~700ns length.

Figure 3 shows measurements for the Au(111) and Re(0001) surfaces recorded in the low-energy range (hν=12-32eV) at beamline U125-2_NIM of BESSY II. The chosen exit slit of 100μm yields a photon bandwidth of only 14meV at hν=21eV. The full time spectrum (referenced to the bunch marker signal) without blanking is shown in Fig. 3(a). The gap in the multibunch train spans from 760 to 900ns. The camshaft pulse (arrow) appears at 840ns and is framed by two additional smaller pulses at the rims of the time gap. After switching on the HF deflector, the signals from the multibunch train and camshaft pulse are separated in the plane of the selector aperture as visible in the real-space images of the field aperture (selector aperture fully open) in Fig. 3(b). The image originating from the multibunch train (b1) is shifted upwards [deflector voltage U(t)=$U_d$ in Fig. 1], whereas the image generated by the camshaft pulse (b2) remains unshifted [deflector voltage U(t)=0]. After driving in the selector aperture by the piezo motors (at the position of the dashed circle), the multibunch train is completely extinguished. Images b1 and b2 have been recorded by adjusting transfer lens 2 to real-space imaging. After checking the proper position of the selector aperture, transfer lens 2 is switched to momentum imaging. For rapid optimization an auxiliary grid is driven into the *backfocal plane* of the objective lens [dashed line left-hand side of Fig. 1]; the image of this 'k-grid' is shown in b3.

The time spectra in Figs. 3(c) for Re(0001) and (d) for Au(111) show just the camshaft pulse plus the two additional smaller pulses, purposely left in the recorded time window for calibration. On the stretched time scale the electron 'pulses' reveal their true nature: They represent complete spectra, here of the Re and Au valence bands. The repetition rate of the camshaft pulse is 1.25MHz (period 800ns); hence the time scale can be strongly stretched such that many time slices are resolved. The Re measurements have been done in standard operation of BESSY II, where the camshaft-pulse current is ~4mA. This yields maximum count rates of >$10^6$ counts per second (cps), being a typical condition for ToF MM in the VUV range. The single-channel DLD can record several $10^6$ cps, a novel multi-line DLD [28] will allow for $10^7$-$10^8$ cps. We notice that DLDs have a fast single-event counting scheme, which is crucial for measurements in the ultrafast time domain [24-26]. The Au measurements were done in low-α multibunch mode with a reduced total beam current of 80mA and a fraction of only 1mA in the camshaft bunch. In this mode with shorter photon pulses (10ps), the corresponding count rate with HF deflector was >$10^5$ cps (photon bandwidth 14meV).

The sequence of momentum patterns in Figs. 3(e-g) shows various sections through the measured 3D $I(E_B,k_x,k_y)$ data arrays recorded in the valence band of Re at photon energies between 12 and 32eV. The full sequence was taken with increments of 1eV (partly 0.5eV) at typical acquisition times of 20 minutes for each photon energy. Rows (e) and (f) show $k_x$-$k_y$ sections at $E_F$ and at higher $E_B$ as given in the panels. Row (g) shows the corresponding $E_B$-vs-$k_x$ sections, revealing the band dispersions.

It is eye-catching that the patterns vary significantly with photon energy. In a MM the k-scale stays constant when varying the kinetic energy. Thanks to the high extractor field, the objective lens is achromatic in a wide range. The microscope records all emitted photoelectrons between $E_{kin}$=0 and $E_F$ in the full half space (2π solid angle). The size of the observed k-region increases with photon energy. Its outer rim in (e,f) is given by the photoemission horizon (emission angle θ=90°). In momentum space the horizon corresponds to a certain radius of the observed $k_\parallel$-region in k-space. The radius at $E_F$ ($E_B$=0) increases from 1.35 to 2.7Å$^{-1}$ if hν is varied between 12 and 32eV.

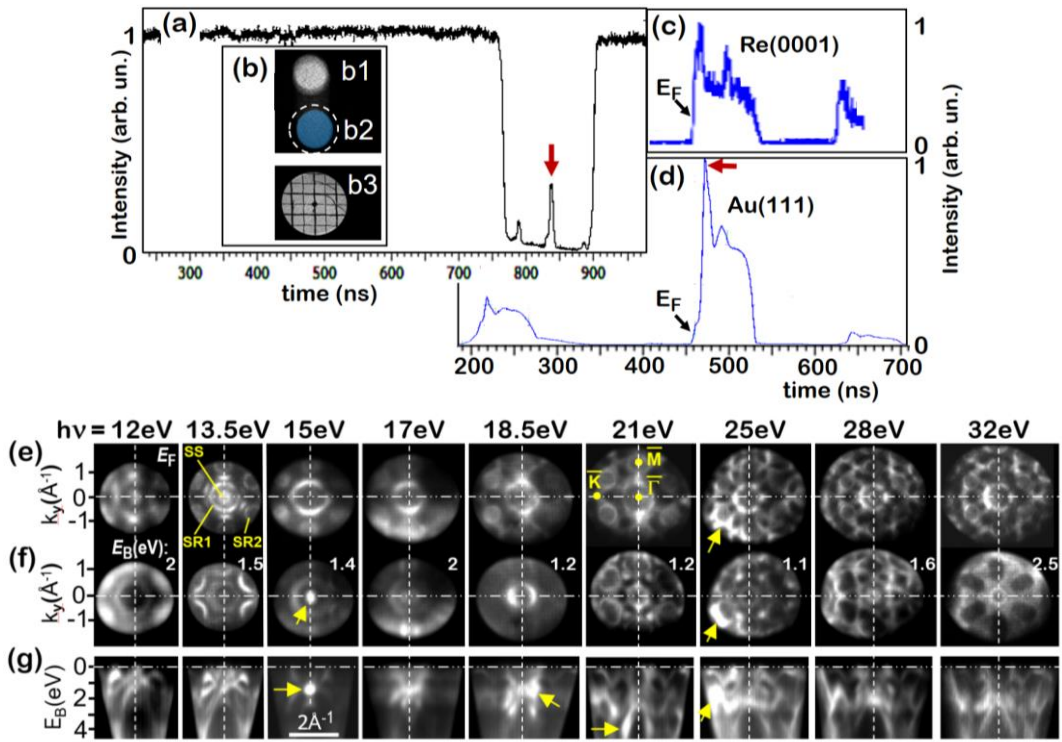

**Figure 3**

Time-of-flight spectra and momentum patterns recorded at BESSY II [U125-2_NIM] in multibunch hybrid-mode (with camshaft pulse). The spectrum for Au(111) without HF deflector (a) is dominated by the multibunch pulse train; the camshaft pulse (arrow) is visible in the centre of the gap at 840ns. After activating the HF deflector as blanking unit (b), the multibunch signal (b1) and camshaft signal (b2) are separated and the former is suppressed by the selector aperture, dashed circle in (b2). The auxiliary grid (b3) is located in the backfocal plane of the objective and serves for precise adjustment of the k-image on the delay-line detector. The time spectra (c,d) for Re(0001) and Au(111) are reduced to the camshaft pulse plus two adjacent small solitaire pulses at the lower and upper edge of the time gap, serving for reference (different filling for Re and Au). On the expanded time scale of (c,d) the 'pulses' appear as full valence-band spectra. Rows (e-g) show momentum sections through data arrays $I(E_B,k_x,k_y)$, recorded for Re(0001) in the chopping mode at photon energies between hν=12 and 32eV as marked on top. (e-g) $k_x$-$k_y$ cuts at $E_F$ (e) and other binding energies as given in the panels (f) and corresponding $E_B$-vs-$k_x$ sections (g) showing the band dispersions. In (e), SS, SR1 and SR2 mark a surface state and two surface resonances; Γ, K and M denote high-symmetry points in the surface Brillouin zone; arrows mark regions of local intensity enhancement by photoelectron diffraction.

Due to photoelectron refraction at the surface barrier, the horizon observed on the vacuum side corresponds to an (inner) escape cone of less than 90°. The $k_\parallel$-momentum is conserved when the photoelectron crosses the surface barrier. The momentum vector $k_\perp$ of photoelectrons inside of the material emitted into the direction of the surface normal varies from 2.40 to 3.32 Å$^{-1}$ for hv=12 to 32eV. In turn, the observed band pattern changes with photon energy, because the photo-transition leads to different final states of the periodic band scheme in 3D k-space. Finally, the patterns exhibit some regions of enhanced intensity due to photoelectron diffraction, intrinsic in the photoemission process, as discussed in [23].

The high-symmetry points $\Gamma$, M, K of the hcp surface Brillouin zone (BZ) as well as a Tamm surface state (SS) and two surface resonances (SR1 and SR2) are denoted in (e). The intensity from the surface state rapidly decreases with increasing residual gas exposure. Because some of the data was not acquired immediately after the flash cleaning, this explains its faint appearance in some of the measurements. In this energy range both surface and bulk bands are observed. The spin-polarized band structure of Re has been studied in Ref. [47], where all details of the band patterns are discussed. Here we just show the suitability of the chopping mode for mapping of the (3D and 2D) BZ. Tomographic scanning of the bulk BZ requires variation of the photoelectron energy $E_{final}$ (inside of the material), which determines the radius of the final-state energy isosphere (for details, see [9]). Given the reciprocal lattice vector $G_{0001}$=1.40Å$^{-1}$ for Re and assuming an effective mass of $m=m_e$ and an inner potential of ~11.5eV, the present photon energy range corresponds to sections through the bulk BZ between 1.7 and 2.4 $G_{0001}$. Since the center ($\Gamma$-point) and the border of the BZ (A-point) lie at 2.0 and 2.5 $G_{0001}$, respectively, the range of hv= 12 – 32eV covers a large range of the bulk BZ. The $\Gamma$-point ($k_z$=2 $G_{0001}$) was identified at hv=18.5eV. For Au(111) the reciprocal lattice vector perpendicular to the surface is $G_{111}$=2.68 Å$^{-1}$. Here the phototransition at a photon energy of 17eV (corresponding to $k_{final}$=2.66 Å$^{-1}$) leads to the $\Gamma$-point of the first repeated BZ ($k_z \approx G_{111}$).

## 2.4 Bandpass pre-selection in the energy domain *versus* chopping in the time domain

In the previous sections we have introduced a method for chopping of the electron pulse train in order to increase the time gap between adjacent electron pulses. Fast beam blanking reduces the fraction of analyzed electron pulses by 1-2 orders of magnitude. However, 3D parallel recording of I($k_x$,$k_y$,$E_B$) data arrays via ToF MM finally gains 2-3 orders of magnitude in recording efficiency. In turn, the net total gain is typically an order of magnitude.

Instead of increasing the time gap at the expense of the amount of analyzed pulses, we can reduce the analyzed energy band, retaining the full number of pulses in the multibunch train. This alternative approach requires the integration of an imaging dispersive energy filter for bandpass pre-selection. Both methods will be used in upcoming setups implementing efficient ToF parallel energy recording at synchrotron-radiation sources (Diamond, PETRA-III) in full multibunch mode.

For the selection of a well-defined electron-energy bandwidth we use a single hemispherical analyzer (HSA) operated as momentum microscope. For optimum recording efficiency, the energy interval transmitted by the HSA is selected such that the desired resolution is reached for a maximum number of time slices resolved by the ToF analyzer behind the exit slit. The width of a single time slice after ToF dispersion defines the energy resolution. The new generation of DLDs provides a time resolution of ≤100ps (<80ps reached [28]). This yields the following numbers of resolved time slices: N=20, 50 or 100 for synchrotron sources with 500, 200 or 100MHz (corresponding time gaps 2, 5, or 10 ns), respectively.

The feasibility of the *dispersive-plus-ToF hybrid mode* on the basis of a large HSA (450mm diameter of the central beam) with ToF section behind the exit slit was proven in Ref. [48]. Here we present quantitative results and give a comparison of the recording efficiency of the HSA with and without the 'ToF booster'. Finally, we compare the expected efficiencies of bunch selection in the time domain by a fast HF deflector and bandpass pre-selection in the energy domain by a dispersive filter.

Figure 4 shows schematic views of the setups for bunch selection (a) and dispersive pre-selection by a HSA (b). The multibunch photoelectron signal, here the Au 4*f* core level (c), is blanked by the HF deflector so that only a single spectrum remains (d), which is repeated with a larger period. This single spectrum is dispersed in the low-energy drift section, leading to the final spectrum (e). In the dispersive+ToF hybrid (b) the energy spectrum (f), here recorded for tungsten, is dispersed in the HSA (operated as MM). The exit slit transmits only a small, well-defined energy interval. This interval is finally dispersed in time by the low-energy ToF drift section, leading to spectrum (g). The desired bandwidth transmitted by the exit slit can be precisely set via the pass energy and HSA slit widths. The subsequent ToF analysis cuts the energy band into N slices (N typically between 20 and 100). Notice that a pure ToF analysis, without the pre-filter, is not possible because full spectra are far too broad to be resolved with the low number of time slices.

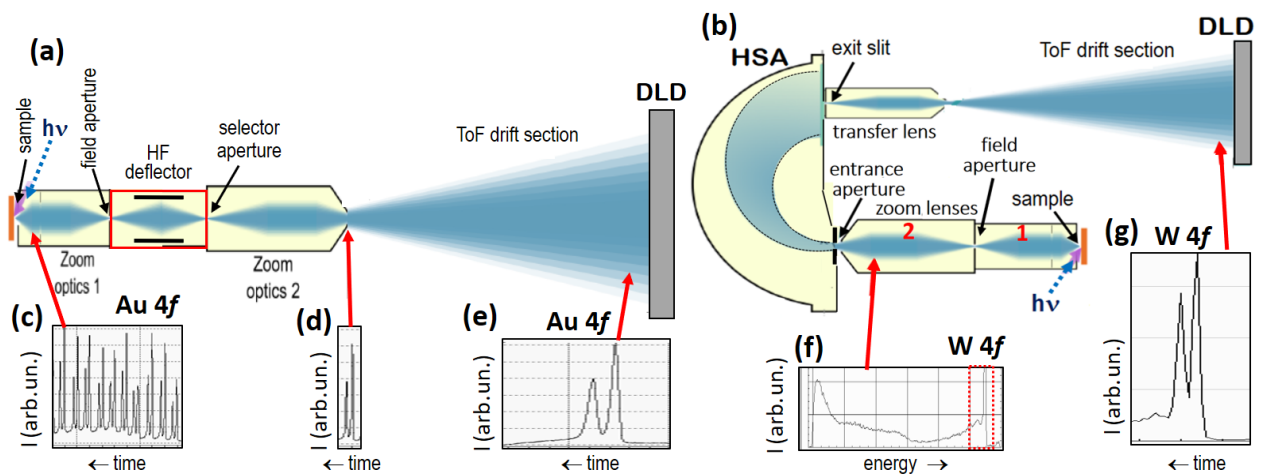

**Figure 4**

Two ways of implementing time-of-flight recording at highly-repetitive photon sources. The first is based on bunch selection in the time domain by fast pulse chopping (a), the second on bandpass selection in the energy domain by a dispersive element (b); electron-optical elements and ray bundles schematic. In (a) the multibunch train of almost-overlapping spectra [Au 4f doublet (c)], is blanked by the HF deflector so that a single spectrum (d) with larger period passes the selector aperture. Spectrum (e) with the desired resolution is gained by ToF dispersion in the low-energy drift section. In the dispersive-plus-ToF hybrid instrument (b), the hemispherical analyzer (HSA) cuts a well-defined bandpass [here W 4f doublet] from the full energy spectrum (f). The transmitted energy band is precisely selected via pass energy and slit widths. This pre-selected energy band is dispersed in the ToF drift section, leading to spectrum (g). In both cases the various lens groups can project either a *momentum pattern* or a *real-space image* on the delay-line detector (DLD). Auxiliary grids (retractable) in the backfocal plane of the objective lens and in the plane of the field aperture allow precise adjustment of the lens optics.

One might ask whether ToF recording bears an advantage, when anyway a dispersive analyzer is used. The answer requires a deeper look into details of dispersive and time-of-flight energy discrimination and will be elucidated in the next section. In fact there is a substantial gain in recording efficiency when using the 'ToF booster' behind the exit slit. The reason lies in the increase of the dimensionality of the data recording scheme from 2D (in the conventional HSA mode as well as in HSA-based MMs) to 3D, when implementing parallel energy recording via ToF detection.

## 2.5 Proof-of-principle of bandpass pre-selection using a UV laser with 80MHz pulse rate

The basic scheme and details of the dispersive-plus-ToF hybrid momentum microscope on the basis of a single-hemisphere HSA are described in [48]. Here we discuss and quantify the gain in recording efficiency by experiments using a high-repetition-rate UV laser (5.8-6.5eV; bandwidth <0.5meV; 80MHz pulse rate [49]). The extractor field collects all photoelectrons in the full solid-angle range of $2\pi$, leading to a paraboloid-shaped $I(k_x,k_y,E_{kin})$ 3D data array. The present single-channel DLD (180ps time resolution) resolves 70 time slices in the 12.5ns gap between adjacent laser pulses. For the evaluation we used N≈50 slices neglecting the time slices close to the rims of the time gap in order to avoid temporal overlap of adjacent pulses.

The results of this pilot study are summarized in Fig. 5. Sample work functions were ~5eV, yielding arrays with depth 1.5eV and diameter ~1.25Å$^{-1}$ at $E_F$. Figs. 5(a-h) show data for the Tamm surface state of Re(0001), (i,j,m-o) for the Shockley state plus quantum-well states on a 3 monolayer film of Au(111) on Re(0001), and for the valence band of 1T-TiTe$_2$ (k,l). Figs. 5(e,f,m-o) show measurements without ToF recording, all other results are recorded using the HSA+ToF hybrid mode.

*Without ToF recording*, the slit width W and pass energy $E_{pass}$ of the HSA have to be set sufficiently small in order to reach the desired energy resolution. Neglecting the $\alpha^2$ term, the theoretical resolution of a HSA with central beam radius $R_0$ is given by $\Delta E_{theo} \approx E_{pass} W/2R_0$. Typical values for the present instrument are W=0.5mm and $E_{pass}$ between 25 and 8eV (yielding $\Delta E_{theo}$ between 28 and 9meV). For high-resolution measurements, W can be reduced down to 0.2mm, yielding $\Delta E_{theo}$=11-2meV for $E_{pass}$=25-4eV.

In the single-hemisphere MM two crucial adjustment conditions must be fulfilled (which is easier than the situation for a double-hemisphere instrument). First, there must be Gaussian (real-space) images in the entrance- and exit-slit planes of the HSA. This is facilitated by shifting an auxiliary grid into the first Gaussian plane ['field aperture' in Fig. 4(b)] and focus its image to the entrance plane of the HSA. Proper setting of the Jost electrodes ensure that this image is transferred 1:1 to the exit plane, an important precondition for good energy and momentum resolution. Fig. 5(a) shows a snapshot of the Gaussian image for a well-adjusted optics: the edge of the exit slit (upper border), the rim of the circular entrance aperture (semi-circle, recorded at $E_F$) and the lines of the auxiliary grid in the field-aperture plane are all in focus. Here the transfer lens behind the HSA exit is adjusted to form a Gaussian image on the detector. Second, the momentum image is optimized shifting a second auxiliary grid into the first reciprocal image plane (backfocal plane of the objective lens). Fig. 5(b) shows a snapshot of the k-image for a well-adjusted optics, where the sharp shadow image of the k-grid is superimposed to the momentum image of a surface state (bright ring). Both conditions need to be fulfilled simultaneously, which requires successive optimization in several steps.

In order to avoid downgrading of the resolution by the $\alpha^2$-term in these first experiments, we limited the entrance angle into the HSA to $\alpha_{max}$=2° by appropriate setting of zoom optics (2 in Fig. 4b). An example recorded with $E_{pass}$=25eV / W=0.2mm ($\Delta E_{theo}$=11meV) is shown in Figs. 5(e,f). In this mode without the ToF booster, the data array of 1.5eV width has been acquired by taking $(k_x,k_y)$ patterns sequentially, varying $E_{kin}$ in steps of 5meV (300 exposures of 20s, total 1.7 hours). The individual $(k_x,k_y)$ patterns are finally concatenated yielding the 3D $I(k_x,k_y,E_{kin})$ data stack.

*When activating the ToF booster*, the pass energy can be increased while retaining the energy resolution. Fig. 5 shows results for $E_{pass}$=200eV (g-j) and 400eV (k,l), both with slits W=1mm, running the ToF drift section at $E_{drift}$=9eV. The as-measured $I(k_x,k_y,\tau)$ data arrays contain the contribution of the transit-time spread of electrons running on different Kepler ellipses in the HSA. The linear time-spread function is directly observable by running the ToF section at high drift energy, where the time dispersion is negligible (here $E_{drift}$=500eV). We observe a linear increase of the transit time with increasing momentum component $k_y$, i.e. with increasing entrance angle $\alpha$ in the dispersive plane, Fig. 5(c). As expected, there is no time spread as function of the entrance angle in the non-dispersive plane, Fig. 5(d); in both cases the entrance angle varies between $\alpha$= -2° and +2°. At $E_{drift}$=9eV the ToF analyzer disperses the transmitted energy band of 880meV width for $E_{pass}$=400eV in time by 8.86 ns. Assuming a time-resolution of 180 ps this results in ~50 resolvable slices of 18meV spacing.

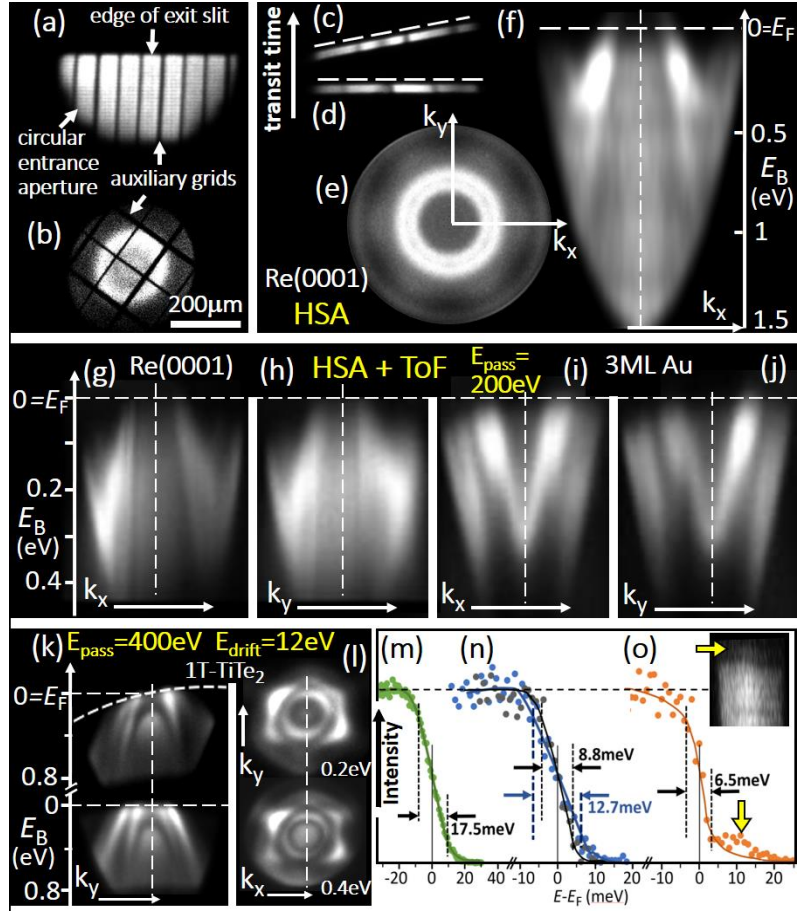

**Figure 5**

Measurements using the bandpass pre-selection approach, recorded with the setup shown in Fig. 4(b) using a narrow-bandwidth UV laser (h$\nu$=6.4eV; bandwidth <0.5meV; 80MHz [49]). (a) Gaussian image showing the superposition of the upper edge of the entrance slit, the circular entrance aperture (half circle) and the grid in the field-aperture plane (vertical lines). (b) Momentum image showing the superposition of the grid in the backfocal plane of the objective lens with a surface state. (c,d) Linear increase of transit time *vs* entrance angle in the dispersive plane (c) and constant behavior in the perpendicular (non-dispersive) plane (d), measured in the HSA+ToF hybrid mode at a drift energy of 500eV. (e,f) Data array recorded using the classical scanning mode of the HSA without ToF analysis: $k_x$-$k_y$ momentum pattern of the Re(0001) Tamm state close to $E_F$ (e) and $E_B$-*vs*-$k_x$ section of the full photoemission paraboloid with diameter ~1.25Å$^{-1}$ at $E_F$ (f). Second row, data recorded in the hybrid mode with $E_{pass}$=200eV and $E_{drift}$=9eV: (g,h) $E_B$-*vs*-$k_{x,y}$ cuts through data arrays of Re(0001) and (i,j), same for 3 ML of Au on Re(0001). (k,l) Data array (width 880meV) for 1T-TiTe$_2$, recorded in the hybrid mode at $E_{pass}$=400eV. (k) $E_B$-*vs*-$k_y$ section as-measured (top) and after numerical elimination of the transit time spread (bottom). (l) $k_x$-$k_y$ sections at $E_B$=0.2 (top) and 0.4eV (bottom). (m-o) Resolution measurements in the conventional scanning mode: intensity profiles of the Fermi edge of a Au film recorded at $E_{pass}$=12eV / W=0.5mm / T=30K (m), 8eV / 0.4mm / 30K (blue) and 15K (black) (n), and 8eV / 0.2mm / 15K (o). Curves in (m,n,o) are to guide the eye; yellow arrows in (o) and the inset mark an intensity artifact slightly above $E_F$.

The combination of the high pass energy of the HSA and the subsequent parallel acquisition of many energy slices via ToF analysis leads to a substantial increase of the recorded intensity. The efficiency gain is defined by the reduction of acquisition time needed to record the same data set with identical statistics, i.e. total number of counts in the I($k_x,k_y,E_{kin}$) array and identical energy resolution. The key factor is the transmission of the HSA, which is defined by the size of entrance slit (restricting the fraction of the electron beam which enters the analyzer) and exit slit (restricting the transmitted energy band). When increasing the pass energy by a factor of N, the bandwidth increases by this factor. Comparing the scanning mode and the HSA+ToF hybrid mode at identical resolution, the gain due to the wider energy band passing the exit slit equals the number of resolved time slices N. The contribution of the entrance aperture is more complex and depends on several factors. The entrance plane hosts a magnified image of the photon spot on the sample. The two-stage zoom optics between sample and entrance aperture [1 and 2 in Fig. 4(b)] allow varying the (total) lateral magnification M in a wide range between M=1 and 40. This wide range allows setting the filling angle of the HSA to any desired value between very small and large filling up to $\alpha_{max}$= +/-7°. In k-imaging mode the $\alpha^2$-term and the non-isochromaticity can be corrected numerically, see detailed discussion in [48].

The behavior of M is governed by the Helmholtz-Lagrange invariant $M\sin\alpha \sqrt{E_{kin}}$=const. Assuming that the full solid-angle range of $2\pi$ above the sample surface is 'squeezed' into a range of $\alpha_{max}$=+/-2.5° inside the HSA, we obtain magnification factors between M=14 (for $E_{pass}$=4eV) and M=1.4 (for $E_{pass}$=400eV). These values reflect the effects of the acceleration from the initial energy (here $E_{kin}$=1.5eV) to the pass energy and the 'angular compression' from 90° to 2.5°. In the general case the magnified image is cropped by the entrance aperture and hence the size of the photon beam plays an important role for the total transmission. For the interplay of $k_{\parallel,max}$, $E_{pass}$, $\alpha_{max}$ and corresponding values of the non-isochromaticity, see eq. (6) and Fig. 4 of ref. [48].

In the small-spot limit (for the present settings at beam diameters <17μm) the magnified image lies completely inside of the entrance aperture for both pass energies. Hence, there is no beam restriction by the entrance aperture and the total gain factor is just the number of resolved time slices, here N≈50. In the limit of large beam diameters (>200μm), the transmitted intensity is proportional to $E_{pass}$. Then a second gain factor N is caused by the entrance aperture. In turn, the total gain factor is N in the small-spot limit and up to $N^2$ in the large-spot limit. Given a diameter of the laser beam of 50μm, the present situation is an intermediate case. Due to the off-normal impact angle of 67.5° the laser beam yields a photon footprint on the sample surface of ~130x50μm$^2$. The magnified image of this photon footprint in the analyzer entrance plane is 180x70μm$^2$ for $E_{pass}$=400eV and 1800x700μm$^2$ for $E_{pass}$=4eV. For $E_{pass}$>60eV the image fits completely into the 0.5mm entrance aperture, whereas for $E_{pass}$=4eV (M=14) only ~16% of the photoelectrons pass through the entrance aperture. For these conditions the total gain factor lies between 300 and 200 for $E_{pass}$ between 4 and 8eV, respectively. We notice that these restrictions do not exist for a pure ToF instrument, because there are no beam-confining apertures except the field aperture defining the ROI.

For an experimental verification, we studied the HSA+ToF hybrid mode at $E_{pass}$=200 and 400eV and slit W=1mm ($\alpha_{max}$<1°). The transmitted bandwidth is 440meV at $E_{pass}$=200eV and 880meV at 400eV, which is sufficient to record the complete occupied part of the Tamm state of Re(0001) and of the Shockley state of Au(111) in one acquisition without scanning. Figs. 5(g-j) show results in terms of $E_B$-vs-$k_{x,y}$ sections through the as-measured I($k_x,k_y,\tau$) data arrays ($\tau$, time-of-flight), without correction of the linear transit time and parabolic $\alpha^2$-term showing up in the $k_y$-sections (h,j). The dashed horizontal line marks the position of the Fermi edge. For 1T-TiTe$_2$, recorded at $E_{pass}$=400eV (k,l), a bandwidth of 880meV is acquired simultaneously. The count rate in this measurement with activated ToF booster was ~3 Mio counts per second (cps); the data array was acquired in 10 minutes. Here we allowed a larger angular filling of $\alpha_{max}\approx2°$, leading to a larger transit-time spread and $\alpha^2$-curvature, visible in (k) in the uncorrected section (top) and corrected by data processing (bottom). An additional curvature results from the very short ToF section (300mm) used for these first experiments. The drift energy was set to $E_{drift}$=9eV, where we measured a temporal dispersion of 22ps per 1meV. The DLD thus resolves time slices of ~18meV each. For comparison, the same data stack was recorded in the scanning mode at $E_{pass}$=16eV using 400μm slit size, which yields a resolution of ~18meV. In this mode the intensity was 20 kcps, i.e. the gain in efficiency was ~150, in fair agreement with the theoretical expectation.

The energy resolution of the HSA without ToF booster has been determined at the Fermi edge of a Au film at various analyzer settings at sample temperatures of 30 and 15K. The measured half widths (16%-84%) of the Fermi-edge cutoff are given in Figs. 5(m-o); settings see figure caption. Deconvolution with the thermal broadening (3kT≈10 and 5meV for 30 and 15K) yields the analyzer resolution $\Delta E_{exp}$, which can be compared with the expected values $\Delta E_{theo}$. For $E_{pass}$=12eV (slits W=0.5mm) $\Delta E_{exp}$=14.4meV is close to $\Delta E_{theo}$=13.3meV (m). For $E_{pass}$=8eV (W=0.4mm) the measured widths for the two temperatures [blue, 30K; black 15K in Fig. 5(n)] yield $\Delta E_{exp}$=7.8 and 7.5meV, in good agreement with $\Delta E_{theo}$=7.1meV. For the smallest slit (W=0.2mm) the value of $\Delta E_{exp}$=4.2meV is still close to the expected value of $\Delta E_{theo}$=3.6meV. The intensity profile Fig. 5(o) reveals a significant artifact slightly above $E_F$ (marked by the yellow arrow). The inset shows that this intensity is not just a high-energy wing but can be located as a spurious additional intensity peak, separated from the Fermi edge. Upon further lowering of $E_{pass}$ this spurious intensity increases and masks the true Fermi cutoff. Close inspection of the Gauss image reveals that this intensity is already visible as a ring-shaped fringe around the image of the entrance aperture in the exit plane. The spurious intensity thus can be traced back to an imaging aberration of the hemisphere itself. Further improvement of the imaging quality might be possible by a modification of the geometry of the Jost plates. This would require ray-tracing calculations using a precise 3D model of the HSA including the details of the fringe field correction.

These first measurements validate that k-microscopy at high resolution is possible for high photon flux in a small spot. The present resolution limit close to 4meV FWHM has been determined for $E_{pass}$=8eV. The instrument allows k-imaging down to $E_{pass}$=4eV. Hence, improvement of resolution by another factor of 2 seems possible, once the imaging aberration of the HSA has been eliminated. In the hybrid mode the energy-resolution limit is defined by the time resolution (180ps for the present DLD). ToF-based k-imaging works well down to $E_{drift}$=6eV providing an energy resolution of 9meV. The combination of the new generation of DLDs (~80ps resolution) with a longer drift tube should yield a resolution limit of $\Delta E_{ToF}$ <5meV. The imaging artifact is absent for the high pass energies used in the hybrid mode.

### 2.6 Displaced 'island-orbit' filling pattern with 5 MHz

Different from the previous methods, which do not affect the storage ring, there is another possibility of implementing ToF into the multibunch operation scheme that was tested at BESSY II. It is based on a special storage ring setting, called TRIBs (transverse resonance island buckets), generating a 2$^{nd}$ stable orbit, wiggling around the main orbit in the equatorial plane [37]. The island orbit can be populated with a single bunch or a few-bunch filling pattern offering reduced repetition rates compared to the multibunch filling on the main orbit. Since the island orbit differs in transverse position and angle from the main orbit, it provides an additional well separated radiation source in bending magnets and undulators, which in turn, can be selected via a pinhole shifted off the equatorial plane, blocking the radiation from the multibunch train. The subsequent monochromator optics corrects for this shift, so that the spot in the exit plane stays the same. Currently, at BESSY II the Top-Up injection in this 'Two-Orbit / TRIBs' mode is under optimization and the bunch current stored on the island orbit is increased, reaching values compared to the Camshaft bunch of the standard hybrid filling (3 - 4 mA). Beyond this application, as bunch separation scheme, offering two different repetition rates simultaneously, the TRIBs setting allows for generating synchrotron radiation with properties not accessible so far. Within a proof-of-principle experiment it was shown that TRIBs enable MHZ-fast helicity flipping of X-rays from an undulator, more than three orders of magnitude faster than state-of-the-art technologies [50]. Especially X-ray circular dichroism (XMCD), one of the main tools to study magnetism, will benefit enormously from this development. The TRIBs /Two Orbit mode is now offered regularly in Top-Up user operation for one week per year at BESSY II.

Due to the high solid-angle acceptance of the U125-2 NIM beamline, it is possible to image the horizontally-offset source point of the ‚island-orbit' in the undulator on the intermediate focus after the first mirror, also horizontally separated from the main beam. By detecting the different pulse rates of the orbits at the experiment, it was possible to insert a horizontal aperture in front of the main beam until only photons with a pulse rate corresponding to that of the "island orbit" of 5 MHz are allowed to pass through the further beamline up to the experiment. This did not require any further optimization of the alignment of the optical elements of the beamline. It shows that it is possible to use the two sources separately at the U125-2 NIM, even in simultaneous operation of the orbits ('twin orbit mode').

The ToF momentum microscope has been used in a test beamtime at the U125-2_NIM using a 4-bunch filling pattern on an island orbit coexisting with the common multibunch pattern on the main orbit. A result measured in this mode is shown in Fig. 6. The sample was a bismuthene monolayer on SiC(0001). This novel material was recently synthesized by epitaxial growth on a hydrogen-etched SiC wafer and studied by STM/STS and photoemission [51]. A band structure similar to graphene but with a large spin-orbit induced total energy gap of ~0.8eV makes bismuthene on SiC a promising candidate for a room-temperature quantum spin Hall system.

Figure 6(a-d) show $k_x$-$k_y$ momentum sections for bismuthene at the valence-band maximum 0.2eV below $E_F$ and at larger $E_B$ as given in the panels. Close inspection of (a) reveals an inner ring of 6 weak spots (corresponding to the 1$^{st}$ BZ) and an outer ring of 6 bright spots (repeated BZ). With increasing $E_B$ the spots open up to threefold-deformed circles with pronounced intensity variations along their periphery. The broadening of the circular patterns at the K points is caused by the spin-orbit induced band splitting [51]. The sketch in (e) shows the bismuthene structure, a honeycomb lattice in perfect registry with the SiC(0001) substrate. The centres of the Bi-hexagons (red) form a ($\sqrt{3}\times\sqrt{3}$)R30° superstructure in on-top position of the topmost Si layer.

The second row (f-j) shows the corresponding $k_x$-$k_y$ patterns for zero-layer graphene on SiC(0001), recorded in a few-bunch filling pattern of BESSY II on the main orbit as described in [43]. The Dirac point lies at the Fermi energy (f) and with increasing $E_B$ the Dirac cones open and show their characteristic triangular shape (g-j). This sample showed a (2x2) superstructure, sketched in the inset in (i): In addition to the well-known six Dirac cones (full triangles) a second ring of Dirac cones with mirrored cross section (dashed triangles) appears. In the sequence (g,h,i) the triangles grow and finally touch each other in (j). Note the striking similarity of the patterns for bismuthene and zero-layer graphene at 1eV below the valence band maximum / Dirac point, (b) and (g), respectively.

The bottom row (k-p) shows $E_B$-vs-$k_{x,y}$ sections along the dashed lines marked in (a) and (f). For bismuthene, (k) shows a cut through a single cone, (l) through two cones of the outer ring (2$^{nd}$ BZ) and one on the inner ring (1$^{st}$ BZ). (m) and (n) show cuts through two cones of the 1$^{st}$ BZ in perpendicular orientations along $k_x$ (m) and $k_y$ (n). For zero-layer graphene, (o) shows a cut through one of the Dirac cones of the bright outer ring and (p) through two cones of the inner (mirrored) ring of the (2x2) superstructure as marked in (f).

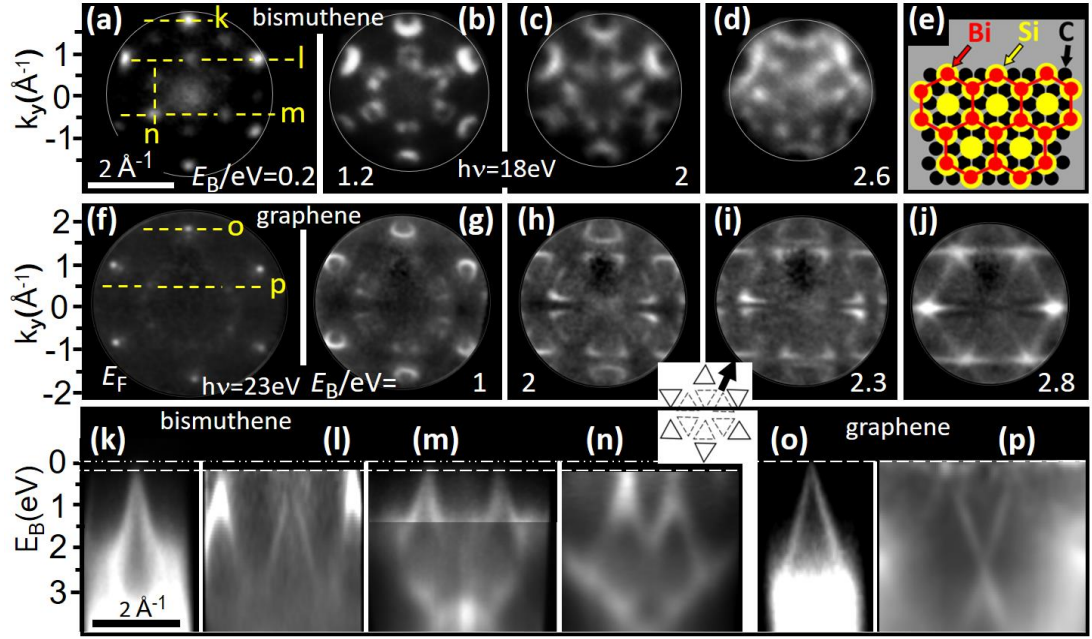

**Figure 6**

Momentum patterns of bismuthene and zero-layer graphene on SiC(0001), recorded at the U125-2_NIM of BESSY II. Results have been recorded for bismuthene in a special 'island-orbit' filling mode of the storage ring, in which 4 bunches travel on an orbit oscillating around the equatorial plane and for graphene in single-bunch mode. (a-d) $k_x$-$k_y$ sections through a data array I($E_B,k_x,k_y$), recorded at h$\nu$=18eV. The valence-band maximum lies 0.2eV below $E_F$ (a); the opening of the cones with increasing binding energy is visible in (b-d). (e) Structure of bismuthene on SiC(0001). (f-g) Analogous results for zero-layer graphene, recorded at h$\nu$=23eV with few-bunch filling of BESSY II. The Dirac point lies at $E_F$ (f) and the opening cones show the well-known triangular cross sections (g-j). For this sample a (2x2) superstructure is present, sketched as dashed triangles in the inset in (i). (k-p) $E_B$-vs-$k_x$ sections showing the linear band dispersions for bismuthene (k-n) and graphene (o,p). (m,n) show the cones of the 1$^{st}$ BZ and (p) shows the 'mirrored cones' of the (2x2) structure.

The bismuthene example proves the usability of the island-orbit mode (coexisting with normal multibunch filling on the main orbit) for ToF experiments. The better k-resolution in the graphene data reflects the higher geometric quality of this film and is not related to the different acquisition modes. For the physics of these systems, see [51,43].

## 3. Conclusions

Three-dimensional photoemission recording schemes in angular- or momentum-resolving ToF instruments exploiting the time structure of synchrotron radiation have proven superior in many respects. Parallel acquisition of many energies via ToF analysis has been shown to minimize the radiation dose for sensitive samples [52], reduce the data-acquisition time in photon-hungry experiments in the soft- [9] and hard-X-ray range [10], and enable full-field XPD [44]. The first few studies indicate that full-field XPD has a huge potential for in-situ structural analysis [19-21], in particular in the ultrafast regime at FELs [26]. Most important, full-field XPD patterns are an indispensable prerequisite for aberration-free bulk band mapping using hard X-rays [11,20-23]. In combination with an imaging spin filter, the 3D-recording architecture enables accessing a full spin texture in reasonable time [47,53-56]. All these experiments used special filling patterns of storage rings (or a mechanical chopper in Ref. [52]), because the short time gap in normal multibunch operation (pulse rates between 100 and 500MHz) is prohibitive for ToF energy discrimination. ToF-based angular- or momentum-resolved photoelectron spectroscopy is superior to 2D recording techniques but suffers from the rare availability of special few-bunch filling patterns of storage rings.

This paper describes two loopholes out of this dilemma. The interplay of desired energy resolution for a certain bandwidth and time resolution of the detector dictates the length of the required time interval. Given present detector resolutions of <100ps, time intervals of >100ns allow resolving about 1000 time slices in the selected energy band of several eV width, which is sufficient for most applications. The first approach is based on a high-frequency deflector unit between two pinholes in the ray path of a momentum microscope. The deflector is phase-locked to the bunchmarker of the storage ring such that it can select a periodic train of electron pulses at a lower repetition rate and correspondingly larger time gap for the ToF analysis. We show examples of 'chopping' the 100MHz photon-pulse train of MAX II to a convenient rate of 5MHz electron pulses and selecting the camshaft pulse (with enhanced bunch charge) at 1.25MHz rate in the hybrid filling pattern of BESSY II with 500MHz multibunch train. A dedicated instrument, selecting each desired pulse rate from the filling pattern (200 MHz) is being developed for PETRA III / PETRA IV.

The second approach retains the full pulse rate but reduces the bandwidth of the electron spectrum being dispersed in the ToF low-energy drift section. A few eV bandwidth can be achieved by a high-pass filter lens in the linear imaging column as described in [43]. But this is not sufficient for pulse rates of $\geq$100MHz. Moreover, high-pass filtering does not allow to select a certain core level because the faster electrons from the valence range or higher-lying core levels would reach the detector as well. For general applications a dispersive element as bandpass filter needs to be implemented in the microscope column. Here we discuss the possibilities opened up with a hemispherical analyzer combined with ToF recording of the energy band passing the exit slit. Fourier-plane imaging in a momentum microscope overcomes the strong variation of the transit time for different Kepler ellipses in the analyzer. Using the new generation of delay-line detectors (~80ps resolution), the ToF section can resolve between N=125 and N=25 time slices for pulse rates between 100 and 500MHz, respectively. The gain in efficiency in the limit of small photon footprints is N. The gain increases to $N^2$ in the limit of large photon footprints, because of the increase in transmission of the analyzer entrance lenses [48]. A prototype instrument will be installed at DIAMOND (beamline I09).

Although they have the same goal, the two methods were developed under different aspects: Pulse-selection in the time domain via electron-optical chopping is an *enabling technique*, opening up a previously-closed avenue to efficient 3D photoemission recording in standard operation modes of storage rings. The hemisphere-plus-ToF hybrid instrument was developed under the aspect of *resolution improvement* by implementing a 'ToF-booster' behind a single-hemisphere momentum microscope. This microscope could be operated without the ToF, although with lower recording efficiency. The present measurements validate that k-microscopy at high resolution is possible for high photon flux in a small spot. A resolution limit of $\Delta E$=4.2meV FWHM has been measured for $E_{pass}$=8eV. This value is limited by the appearance of an artifact in the image of the entrance slit in the exit plane of the HSA. The present setup allows k-imaging down to $E_{pass}$=4eV. Hence a further improvement of resolution by a factor of 2 seems possible, after improvement of the imaging quality of the HSA. In the hybrid mode at high pass energies the imaging artifact is absent and the energy-resolution limit is defined only by the time resolution. k-imaging works well down to $E_{drift}$=6eV in the ToF section resulting in an energy resolution of 9meV. For a newer generation of DLDs with 80ps resolution and a longer drift tube we expect a resolution of $\Delta E_{ToF}$ <5meV.

Bandwidth pre-selection in the energy domain does not require a k-imaging hemisphere. A resolution of 30meV (typical photon bandwidth of a soft-X-ray beamline) at a 100MHz storage ring and a DLD with 80ps resolution (N=125) would require narrowing the energy bandwidth to 3.8eV. This can be done by a much simpler dispersive bandpass filter on the basis of a *multipole-deflector doublet* in an almost straight microscope column. Such a filter is under development.

**Acknowledgements**

We thank the staff of MAX II (Lund, Sweden) and BESSY II at Helmholtz-Zentrum Berlin (Germany) for excellent support during several beamtimes. Further thanks are due to Ralph Claessen and his group (University of Würzburg, Germany) for good technical cooperation and the provision of the bismuthene sample, and to Thomas Allison (Stony Brook University, USA) for fruitful discussions on ToF MM at high pulse rates and a critical reading. We also thank Thorsten Kampen, Oliver Schaff and Sven Maehl (SPECS GmbH, Berlin) for discussions on details of the hemispherical analyzer and Tobias Grunske, Tatjana Kauerhof and colleagues (APE GmbH, Berlin) for the loan of the UV laser and excellent support. Funding by BMBF (05K16UMB and 05K19UM1) and Deutsche Forschungsgemeinschaft DFG (German Research Foundation)—TRR 173–268565370 (project A02) and TRR 288–422213477 (project B04) is gratefully acknowledged.